\documentclass[aps,prl,a4paper,twocolumn,groupedaddress,english,longbibliography]{revtex4-1}
\usepackage{amsmath,amssymb}
\usepackage{graphicx,epsfig,sidecap,wasysym}
\usepackage{color,epstopdf}
\usepackage[colorlinks,bookmarks=false,citecolor=blue,linkcolor=blue,urlcolor=blue]{hyperref}

\def\be{\begin{equation}}
\def\ee{\end{equation}}

\def\bea{\begin{eqnarray}}
\def\eea{\end{eqnarray}}

\newcommand{\ket}[1]{|\,#1\,\rangle}

\begin{document}
\title{Multifractal Scalings across the Many-Body Localization Transition}
\author{Nicolas Mac\'e}
\email{nicolas.mace@irsamc.ups-tlse.fr}
\affiliation{Laboratoire de Physique Th\'eorique, IRSAMC, Universit\'e de Toulouse, CNRS, UPS, France}
\author{Fabien Alet}
\email{fabien.alet@irsamc.ups-tlse.fr}
\affiliation{Laboratoire de Physique Th\'eorique, IRSAMC, Universit\'e de Toulouse, CNRS, UPS, France}
\author{Nicolas Laflorencie}
\email{nicolas.laflorencie@irsamc.ups-tlse.fr}
\affiliation{Laboratoire de Physique Th\'eorique, IRSAMC, Universit\'e de Toulouse, CNRS, UPS, France}
\begin{abstract}
In contrast with Anderson localization where a genuine localization is observed in real space, the many-body localization (MBL) problem is much less understood in the Hilbert space, support of the eigenstates. In this work, using exact diagonalization techniques we address the ergodicity properties in the underlying ${\cal{ N}}$-dimensional complex networks spanned by various computational bases for up to $L=24$ spin-1/2 particles ({\it{i.e.}} Hilbert space of size ${\cal{N}}\simeq 2.7\,10^6$). We report fully ergodic eigenstates in the delocalized phase (irrespective of the computational basis), while the MBL regime features a generically (basis-dependent) multifractal behavior, delocalized but non-ergodic. The MBL transition is signaled by a non-universal jump of the multifractal dimensions.
\end{abstract}
\maketitle
\noindent{\bf Introduction---} 
Anderson localization (AL)~\cite{anderson_absence_1953,evers_anderson_2008} is a fundamental phenomenon where transport is hindered by disorder in quantum systems composed of free particles.
When interactions between the particles is added, localization tends to be prohibited and thermalization is favored.
 The eigenstate thermalization hypothesis (ETH)~\cite{deutsch_quantum_1991,srednicki_chaos_1994}, which provides a theoretical understanding of how thermal equilibrium is encoded at the level of individual many-body eigenfunctions, applies generically to most interacting systems (with the important exception of integrable models). 
 It is thus quite remarkable that disordered quantum interacting systems can still show absence of transport and thermalization in the form of many-body localization (MBL)~\cite{gornyi_interacting_2005,basko_metalinsulator_2006}, a topic that has attracted considerable interest recently (see reviews Ref.~\onlinecite{nandkishore_many-body_2015,abanin_recent_2017,alet_many_2018,abanin_ergodicity_2018}). 
 Besides arrested transport, MBL features several striking properties, {\it{e.g.}}  anomalously small (area-law) entanglement~\cite{bauer_area_2013}, emergent integrability and Poisson spectral statistics~\cite{imbrie_local_2017,rademaker_many-body_2017}, very slow (logarithmic) entanglement spreading~\cite{znidaric_many-body_2008,bardarson_unbounded_2012,serbyn_universal_2013}, some of which expose the differences between MBL and AL. 
 Numerical simulations of 1d lattice models~\cite{pal_many-body_2010,luitz_many-body_2015}, analytical rigorous proofs~\cite{imbrie_diagonalization_2016}, and effective localized-bits ({\it l}-bits) models~\cite{huse_phenomenology_2014,serbyn_local_2013,imbrie_local_2017,rademaker_many-body_2017} have been instrumental in understanding these properties of the MBL phase. 
\begin{figure}[hb]
\includegraphics[width=.9\columnwidth]{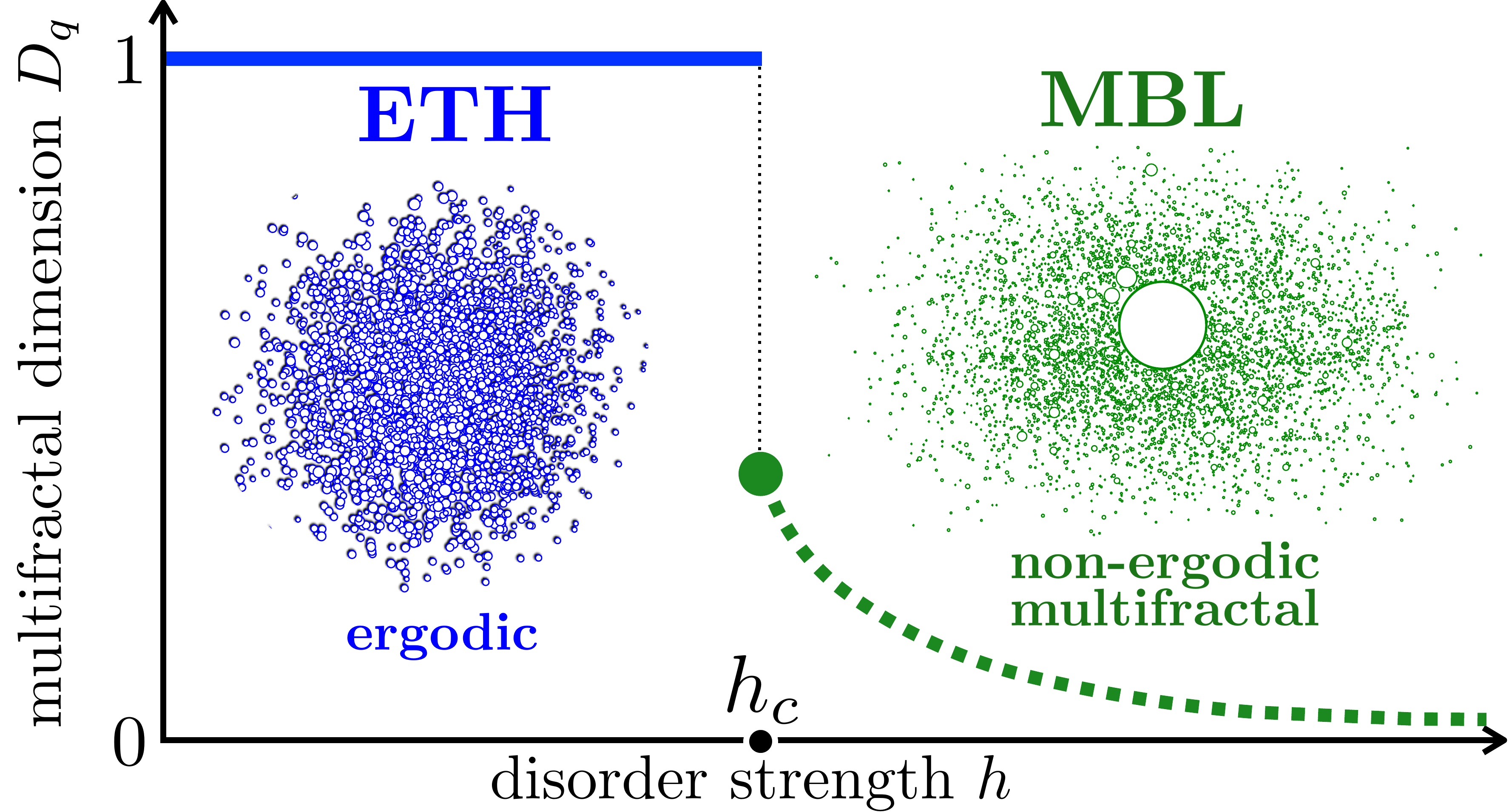}
\caption{Schematic picture of the multifractal properties of eigenstates in Fock and spin configuration bases across the MBL transition for Eq.~\eqref{eq:Hs}. Two typical eigenstates of Eq.~\eqref{eq:Hs} on a small $L=14$ system for ETH ($h=0.5$, blue) and MBL ($h=10$, green) regimes are graphically represented, with circle sizes proportional to $|\psi_\alpha|^2$ in the spin configuration basis.}
\label{fig:1}
\end{figure}

However, several important questions remain open regarding the MBL-ETH transition and the nature of eigenstates. This contrasts with the much-better understood case of AL where the spatial extension of {\it single-particle} orbitals is known to display {\it multifractality} at criticality on finite-dimensional lattices, a behavior which characterizes the transition and its underlying field-theoretical description~\cite{evers_anderson_2008}. In the MBL context on the other hand, the behavior of {\it many-body} eigenstates in the Hilbert space is not firmly established~\cite{luca_ergodicity_2013,luitz_many-body_2015,pino_nonergodic_2016,torres-herrera_extended_2017,luitz_ergodic_2017,pino_multifractal_2017,buijsman2018}. {{We aim at addressing three open issues:}}  (i) the possible existence of an intermediate non-ergodic delocalized phase in the ETH phase~\cite{pino_nonergodic_2016,torres-herrera_extended_2017,pino_multifractal_2017}; (ii) the wavefunctions properties in the MBL regime sometimes described as Fock-space localized~\cite{pino_multifractal_2017,kravtsov_non-ergodic_2018}, {{while perturbative~\cite{gornyi_spectral_2017,tikhonov_many-body_2018} and numerical~\cite{luitz_many-body_2015} results indicate multifractality}}; (iii) {{a consistent description of the critical regime.}}

The central quantities of interest in our work are the participation entropies (PE) $S_q$, derived from the $q^{\rm th}$ moments of a wavefunction $|\Psi\rangle$ expressed in a given basis:
\be
|\Psi\rangle=\sum_{\alpha=1}^{\cal N} \psi_\alpha|\alpha\rangle\,\,\,{\rm{and}}\,\,\,S_q=\frac{1}{1-q}\ln\left(\sum_{\alpha=1}^{\cal N} |\psi_\alpha|^{2q}\right).
\label{eq:Sq}
\ee
In the Shannon limit $S_1=-\sum_{\alpha} |\psi_\alpha|^{2}\ln |\psi_\alpha|^{2}$, while the case $q=2$ recovers the usual inverse participation (IPR)~\cite{visscher_localization_1972} with $S_2=-\ln \left({\rm IPR}\right)$. 
Let us first describe the possible leading asymptotic behaviors for $S_q$ with the support size $\cal N$. 
For a perfectly {\it{delocalized}} state $S_q=\ln {\cal N}$.
In contrast, if a state is {\it{localized}} on a finite set, one gets $S_q={\rm{constant}}$, as observed for AL. In an intermediate situation, wavefunctions are extended but non-ergodic, with $S_q=D_q\ln {\cal N}$ ($D_q< 1$ are $q$-dependent multifractal dimensions).

In this Letter, we inspect {\it many-body} eigenstates in the Hilbert space by studying $D_q$ across the ETH-MBL transition for model  Eq.~\eqref{eq:Hs}. 
Building on large-scale numerics for the most relevant basis sets $\{|\alpha\rangle\}$, {{we provide a complete scaling analysis describing both phases and the critical point, which captures finite-size effects.}} We conclude for full ergodicity ($D_q=1$) for ETH, generic multifractality in the entire MBL regime, and a non-universal jump of $D_q$ at the transition, as sketched in Fig.~\ref{fig:1}~\cite{Note1}.

\noindent{\bf MBL as a complex network Anderson problem---}
We focus on the random-field XXZ $S=\frac{1}{2}$ chain model
\be
{\cal H}=\sum_{i=1}^{L}\Bigl[\Delta S_i^z S_{i+1}^{z}-h_iS_i^z+\frac{1}{2}\left(S_i^+ S_{i+1}^{-}+S_i^- S_{i+1}^{+}\right)\Bigr],
\label{eq:Hs}
\ee
{{with periodic boundary conditions}} and $h_i$ randomly drawn from a uniform distribution $[-h,h]$. Eq.~\eqref{eq:Hs} is equivalent~\cite{sm}~
 to interacting spinless fermions in a random potential.
This system has been intensively studied~\cite{znidaric_many-body_2008,pal_many-body_2010,bauer_area_2013,luca_ergodicity_2013,luitz_many-body_2015,serbyn_criterion_2015,logan_many-body_2018} and its phase diagram is well-known for the case $\Delta=1$ with a critical disorder {{estimated to be}} $h_c=3.7(2)$ in the middle of the many-body spectrum $\{E\}$ such that $\epsilon=(E-E_{\rm min})/(E_{\rm max}-E_{\rm min})=0.5$. This Hamiltonian can be recast as a single particle Anderson problem of the general form
\be
{\cal H}=\sum_{\alpha}\mu_\alpha|\alpha\rangle\langle \alpha| +\sum_{\langle \alpha\beta\rangle }t_{\alpha\beta}|\alpha\rangle\langle \beta|,
\label{eq:Hc}
\ee
in a given basis $\{|\alpha\rangle\}$.
Of course, the localization properties measured by the PE depend crucially on the choice of $\{|\alpha\rangle\}$. We focus on two bases: spin configurations $\{|\alpha\rangle\}_S$ and Fock basis $\{|\alpha\rangle\}_F$, which we argue are the most relevant for the model Eq.~\eqref{eq:Hs}: (i) both bases diagonalize $\cal H$ in specific limits where localization is well-understood (the non-interacting limit $\Delta=0$ for $\{|\alpha\rangle\}_F$, the limit $h\rightarrow \infty $ for $\{|\alpha\rangle\}_S$) and are thus used as a starting point for the {\it l}-bits construction or efficient numerical simulations of MBL;  {(ii)} they implement the U(1) conservation rule (particle number or magnetization conservation) of the model; 
(iii) ${\cal H}$ is {\it sparse} in both bases: an ingredient which eases  numerical diagonalization and may favor ergodicity breaking.

{\it{a) Spin configuration basis.}}
The basis $\{|\alpha\rangle\}_S$ uses the local projection of $S_i^z$, {\it{i.e.}} $\ket{\alpha}=\ket{\hskip -0.15cm \uparrow\downarrow\uparrow\ldots}$. We restrict the study to the zero magnetization sector $\sum_i S_i^z =0$ (half-filling for fermions) of dimension ${\cal N}={{L}\choose{L/2}}\simeq 2^L/\sqrt{L}$. In the spin configuration basis, Eq.~\eqref{eq:Hs} becomes a hopping problem Eq.~\eqref{eq:Hc} with disordered on-site energies $\mu_\alpha=\langle\alpha|\sum_i\Delta S_i^z S_{i+1}^{z}-h_iS_i^z|\alpha\rangle$, and constant hopping terms $t_{\alpha\beta}=1/2$ (allowing tunnelling between neighboring states $\langle \alpha\beta\rangle$ connected by spin-flip terms of Eq.~\eqref{eq:Hs}). As the site-dependent connectivity ${\overline{z}}\approx L/2$ grows faster with system size $L$ than the average on-site disorder strength $\sigma_\mu
\sim h\sqrt{L}$, general arguments would prohibit genuine AL in this complex network (note however the strong correlations of the potential between neighboring sites).

\begin{figure}[b]
\includegraphics[clip,width=\columnwidth]{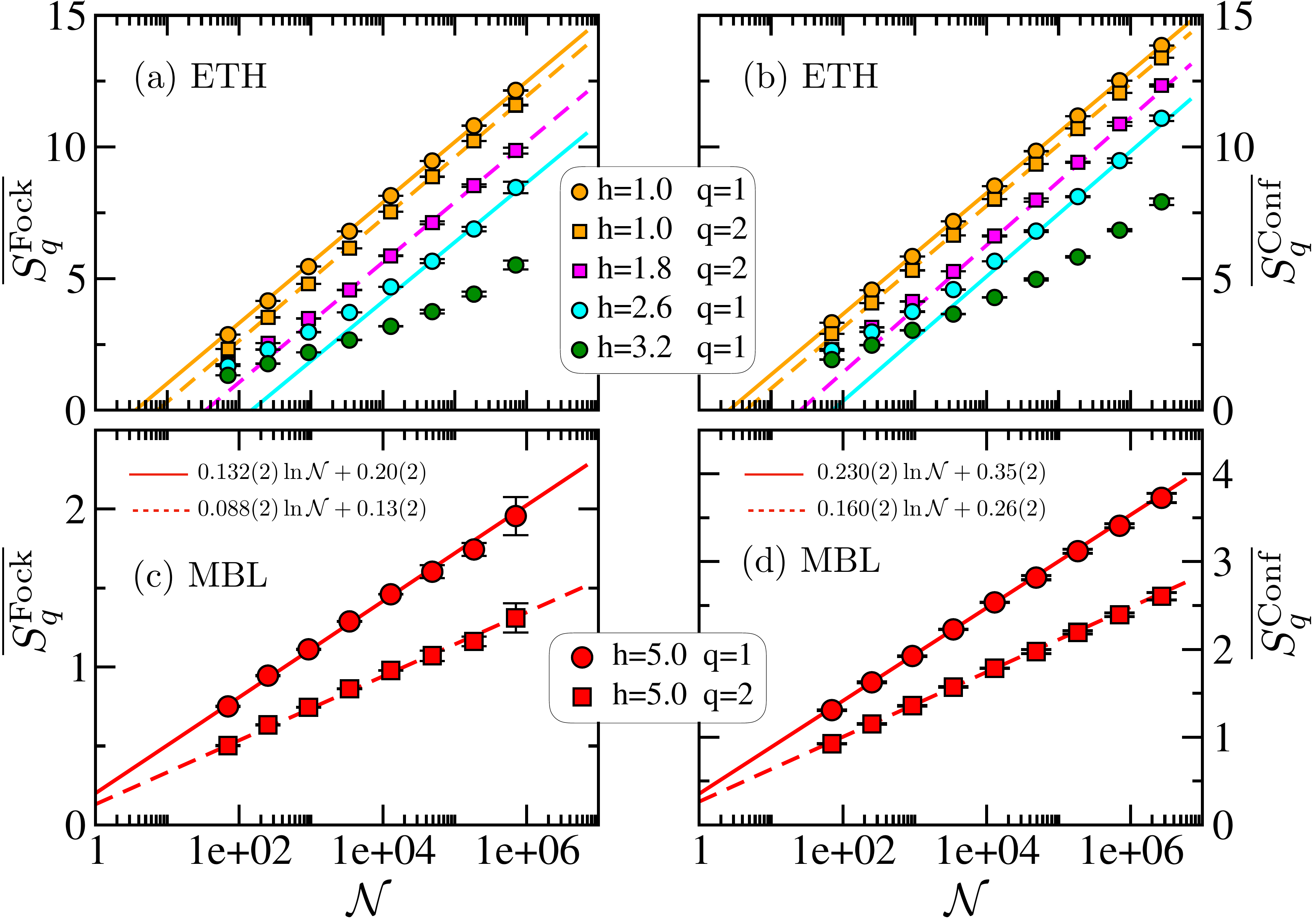}
\caption{Finite size scaling of the disorder average PEs ${\overline{S_q}}$ shown for both Fock (left: a,c) and configuration spaces (right: b,d) and a few representative values of disorder strength $h$ in  both ETH (a,b) and MBL (c,d) regimes for $q=1,2$. Lines are of the form $D_q\ln {\cal N}+ b_q$ with $D_q=1$ and $b_q<0$ in the ETH regime {{(guides to the eye)}}, while $D_q<1$ and $b_q>0$ in the MBL regime {{(fits)}}.}
\label{fig:2}
\end{figure}

{\it{b) Fock basis.}}
 $\{|\alpha\rangle\}_F$ are many-body states built from non-interacting localized orbitals which diagonalize the free-fermion part ${\cal H}-\Delta\sum_i S_i^z S_{i+1}^{z}$ of Eq.~\eqref{eq:Hs}.
 On-site potentials $\mu_\alpha$ are the sum of the non-interacting orbital energies corrected by the Hartree-Fock term, and off-diagonal hoppings $t_{\alpha\beta}$ are built from the interaction terms~\cite{bauer_area_2013,prelovsek_reduced-basis_2018,sm}.  Viewing MBL as an Anderson problem on graphs defined by these Fock states has been promoted in very early works~\cite{altshuler_quasiparticle_1997,gornyi_interacting_2005,basko_metalinsulator_2006}. Nevertheless, the hopping problem expressed in $\{|\alpha\rangle\}_F$ is qualitatively different than from in $\{|\alpha\rangle\}_S$. While diagonal terms have similar behaviors, there is a much larger number of non-zero matrix elements $z_{\alpha\beta}$ between Fock states which is constant over the graph: $z_{\alpha\beta}=\frac{1}{4}\left[\frac{L}{2}\left(\frac{L}{2}-1\right)\right]^2+\frac{L^2}{4}$ (in the $\sum_i S_i^z=0$ sector). Moreover, hoppings $t_{\alpha\beta}$ are not constant but random, both in sign and magnitude~\cite{sm}.

\begin{figure*}[ht!]
\centering
\includegraphics[clip,width=1.9\columnwidth]{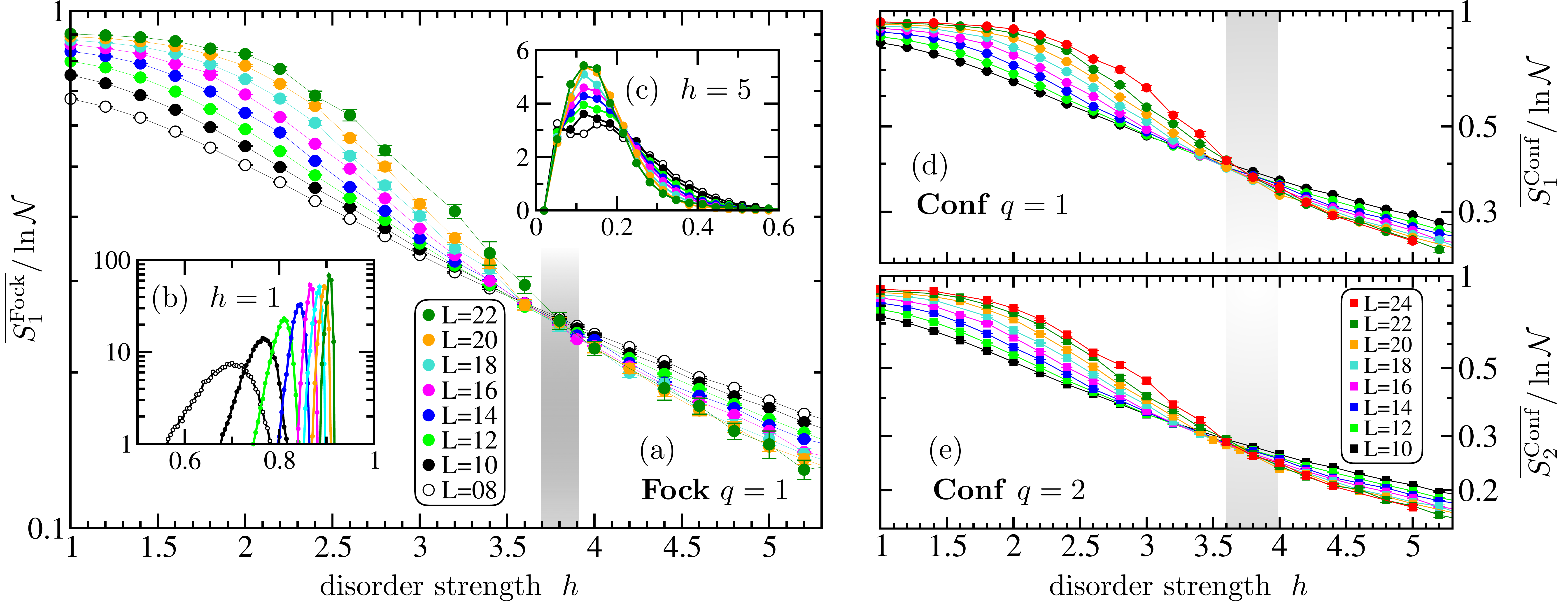}
\caption{Scaled PE ${\overline{S_q}}/\ln \cal N$ plotted against disorder $h$ in both Fock (a) and configuration spaces (d,e). Crossing points signal a sign change of the subleading correction $b_q$ in the PE scalings $D_q\ln {\cal N}+b_q$ which occurs in the vicinity of the ETH-MBL transition point $h_c\sim 3.8$ (gray shaded region). Insets (b,c) show histograms $P(S_{1}^{\rm Fock}/\ln {\cal N})$ of PE in the Fock basis.}
\label{fig:3}
\end{figure*}
\vskip 0.15cm
\noindent{\bf Participation entropies: large scale numerics---} Eigenstates of Eq.~\eqref{eq:Hs} are extracted from full and shift-invert subset~\cite{pietracaprina_shift_2018} diagonalization, focusing on high-energy states ($\epsilon=0.5$) on various system sizes, ranging from $L=8$ spins (with ${\cal N}=70$) up to $L=24$ (${\cal N} = 2\,705\,432$). Disorder average is performed over many independent samples, typically tens of thousand for the smallest sizes $L\le 16$, and several hundreds for the largest samples $18\le L\le 24$~\cite{noteED}. We fix $\Delta=1$.

We first discuss the scaling with Hilbert space size $\cal N$ of the disorder-average PE ${\overline{S_q}}$ defined by Eq.~\eqref{eq:Sq} and shown in Fig.~\ref{fig:2} for both Fock and configuration spaces for a few representative values of the disorder.
In the ETH phase (Fig.~\ref{fig:2} a,b), low disorder data follow a purely ergodic scaling of the form $S_q=\ln {\cal N}+b_q$ with $b_q<0$ for both Fock and configuration spaces. Upon increasing $h$ a curvature develops, indicating that an asymptotic scaling regime might be eventually reached for larger $\cal N$, as exemplified by $h=2.6$ and $h=3.2$ data. In the MBL regime (Fig.~\ref{fig:2} (c,d) with $q=1,2$) at $h=5$ both bases exhibit a delocalized behavior $S_q=D_q\ln {\cal N}+b_q$ with a $q$- and basis-dependent multifractal dimension $D_q<1$ and a  correction $b_q>0$. 

\begin{figure*}[ht!]
\includegraphics[clip,width=2\columnwidth]{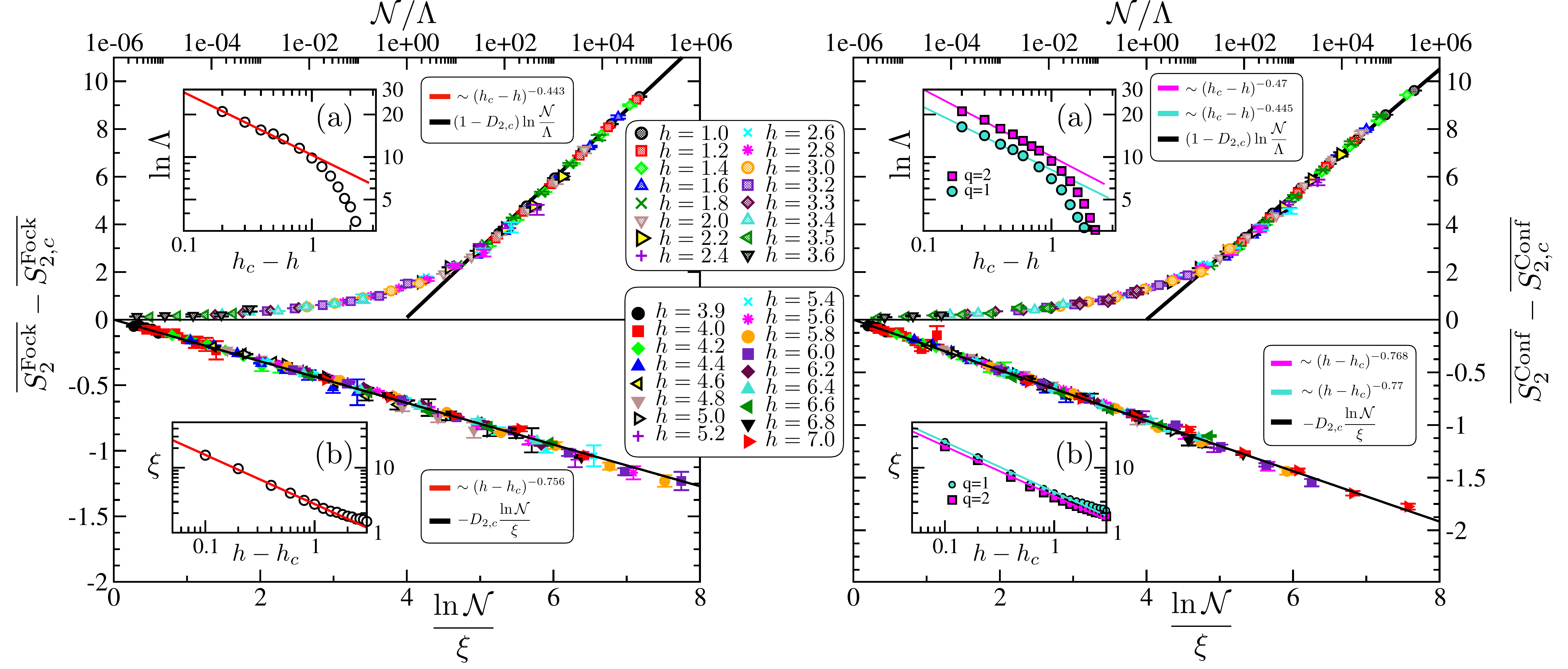}
\caption{Scaling curves for the PE following Eq.~\eqref{eq:scaling} across the full ETH-MBL  regimes with varying disorder strengths (colored symbols). Data for $q=2$ are displayed for both Fock (Left) and configuration (Right) spaces, after subtracting the critical PE data ${\overline{S_{2,c}}}$ taken at $h=3.8$. For $h<h_c$ (top panels) a volumic scaling ${\cal{G}}_{\rm vol}({\cal N}/\Lambda)$ gives the best data collapse. The non-ergodicity volumes $\Lambda$ plotted in insets (a) show similar (universal) divergences $\ln \Lambda\sim (h_c-h)^{-a}$  with $a\sim 0.45$ for Fock ($q=2$) and configuration ($q=1$ and $q=2$) spaces, as indicated on the plot. Bottom panels show data collapses in the MBL regime for $h> h_c$ following a linear scaling ${\cal{G}}_{\rm lin}(\ln{\cal N}/\xi)$. The length scale $\xi$ plotted in insets (b) diverges as $(h-h_c)^{-a'}$ with $a'\sim 0.76$. Note that small deviations to scaling Eq.~\eqref{eq:MBL} are expected due to non-universal subleading constant corrections $b_q>0$, but are difficult to observe in this strong disorder regime.}
\label{fig:4}
\end{figure*}

Very interestingly, $b_q$ changes sign (negative for ETH and positive for MBL) as seen in Fig.~\ref{fig:3} where a crossing point in ${\overline{S_q}}/\ln \cal N$ appears in the vicinity of the critical disorder strength $h_c\sim 3.8$. Crossings are equally observed for Fock and configuration spaces in Fig.~\ref{fig:3} (a,d,e). This effect is also clearly visible from the distributions $P\left(S_q/\ln\cal N\right)$ shown in Figs.~\ref{fig:3} (b,c) where we observe strikingly distinct behaviours on both sides of the transition, with qualitatively different finite size effects and opposite skewness. As $S_q/\ln\cal N$ is restricted to the interval $[0,1]$, the distributions at small disorder slowly converge (from below) towards unity with system size. Conversely, in the MBL regime, rare events tails extend above the average while distributions shrink with system size {{(with a variance vanishing as a power-law~\cite{sm})}}, thus leading to positive finite size corrections. As argued below, the negative constant correction $b_q$ in the ETH phase can be physically related to a non ergodicity volume $\Lambda_q=\exp(-b_q)$. In contrast,  the positive value of $b_q$ in the MBL regime seems to be rooted in rare events showing up in positive skewness distributions.
\vskip 0.15cm
\noindent{\bf Scaling analysis---} We turn to a finite size scaling analysis of PE across the MBL transition. Fig.~\ref{fig:4} shows results for $q=2$ in the two basis where very good collapses can be obtained using the following ansatz~\cite{note_collapse}, inspired by a recent investigation~\cite{garcia-mata_scaling_2017} of the Anderson problem on random graphs:
\begin{eqnarray}
{\overline{S_q}}({\cal N})-{\overline{S_{q,c}}}({\cal N}) =
  \begin{cases}
   {\cal G}_{\rm vol}\left(\frac{\cal N}{\Lambda_q}\right) & \text{if } h< h_c \\
{\cal G}_{\rm lin}\left(\frac{\ln \cal N}{\xi_q}\right)       & \text{if } h> h_c.
  \end{cases}
  \label{eq:scaling}
  \end{eqnarray}
%

\noindent Note that we first subtracted ${\overline{S_{q,c}}}({\cal N})$ the critical PE at $h_c=3.8$~\cite{Note_hc}. In the ETH regime (top panels), a volumic scaling ${\cal G}_{\rm vol}({\cal N}/\Lambda)$ yields an excellent description of the crossover with $\Lambda$ interpreted as a non-ergodicity volume~\cite{garcia-mata_scaling_2017}. When the Hilbert space dimension ${\cal N}\gg \Lambda$, full ergodicity is recovered with $S_q=\ln {\cal N}-(1-D_{q,c})\ln \Lambda$,  $D_{q,c}$ being the multifractal dimension at the transition, while the critical behavior $S_q\to S_{q,c}$  is recovered below the non-ergodicity volume  ${\cal N}\ll \Lambda$. This volumic scaling  is found to be universal as it occurs for both Fock and spin basis, with a non-ergodicity volume diverging very fast as the transition is approached: $\ln \Lambda\sim (h_c-h)^{-a}$ (insets (a) in Fig.~\ref{fig:4}), with $a\approx 0.45$~\cite{note_PL}. An analogous volume law with $a= 0.5$ was reported~\cite{tikhonov_anderson_2016,garcia-mata_scaling_2017} for the Anderson problem on random regular graphs, in agreement with analytical predictions in infinite dimensions~\cite{mirlin_universality_1991,fyodorov_localization_1991,fyodorov_novel_1992}.

Contrasting with ETH, data in the MBL regime rather follow a linear scaling function ${\cal{G}}_{\rm lin}(\ln{\cal{N}}/\xi)$ with a universal form
\be
S_{q,{\rm MBL}}-S_{q,c}=-D_{q,c}\frac{\ln {\cal N}}{\xi},
\label{eq:MBL}
\ee
again yielding a very good data collapse (bottom panels of Fig.~\ref{fig:4}).
The length scale $\xi$, extracted from the best collapse, is shown in panels (b) of Fig.~\ref{fig:4} as a function of the distance to criticality $h-h_c$, where again $h_c=3.8$ has been fixed. In Eq.~\eqref{eq:MBL}, $\xi$ lies in the range $(1,+\infty)$ which guarantees $S_{q,{\rm MBL}}$ to remain positive in the limit $\ln {\cal N}\gg \xi$ where the leading term follows $S_q=D_{q,c}\left(1-1/\xi\right)\ln\cal N$. In the other limit $\ln {\cal N}\ll \xi$, one retrieves the critical scaling. Close to criticality, we observe a divergence of the length scale $\xi\propto (h-h_c)^{-a'}$, with $a'\approx 0.76$.

It is crucial to make a distinction between random graphs which display  AL at large disorder {{(see for instance Refs.~\cite{de_luca_anderson_2014,tikhonov_anderson_2016,garcia-mata_scaling_2017})}}, and the present many-body problem where instead, multifractality is present in the entire MBL phase $\forall h\ge h_c$. 
We clearly observe~\cite{sm} $D_{q,{\rm MBL}}(h\gg h_c) \propto 1/h$, thus vanishing only in the infinite disorder limit, in agreement with recent results~\cite{tikhonov_many-body_2018}. This behavior can be accounted for by simple strong disorder arguments~\cite{sm}. Numerical data further suggest that this form persists almost down to the critical point such that $D_{q,{\rm MBL}}(h) \simeq D_{q,c}\frac{h_c}{h}$.

\vskip 0.15cm
\noindent{\bf Discussion---} Finite-size effects might affect the estimates of MBL critical point and exponents~\cite{chandran_finite_2015,PhysRevLett.119.075702}. However our scaling analysis carefully accounts for them (Fig.~4) and strongly supports the following scenario across the ETH-MBL transition for the standard model Eq.~\eqref{eq:Hs}. (i) Eigenstates are generically ergodic, with no multifractality over the entire ETH phase ($D_q=1$) thus removing the possibility for an intermediate non-ergodic state with $D_q<1$ in the model Eq.~\eqref{eq:Hs}. (ii) The entire MBL regime harbors delocalized non-ergodic eigenstates with disorder and basis dependent multifractal dimensions, only vanishing in the infinite disorder limit. (iii) The ETH-MBL transition point belongs to the same non-ergodic regime, with a jump of the multifractal dimension $D_q<1$ at the transition, in agreement with recent predictions for a non-thermal critical point~\cite{khemani_critical_2017,dumitrescu_scaling_2017,thiery_many-body_2018,thiery_microscopically_2017,dumitrescu_kosterlitz-thouless_2019}. ETH-MBL criticality displays two radically different scalings across the transition for eigenstates extension through the Hilbert space: a volumic law ${\cal{G}}_{\rm vol}({\cal N}/\Lambda)$ for ETH with an emergent non-ergodicity volume $\Lambda$, diverging at the transition, and a linear scaling ${\cal G}_{\rm lin}(\ln {\cal N}/\xi)$ in the MBL regime. {{This is clearly compatible with a second-order critical point}}, but with an unsymmetrical nature, as also recently highlighted in Refs.~\cite{dumitrescu_scaling_2017,thiery_microscopically_2017,khemani_critical_2017,goremykina_analytically_2019}. To which extent $\xi$ is related to a localization length in real space remains to be understood. 
{{{We do not see signs of a putative second MBL phase~\cite{dumitrescu_kosterlitz-thouless_2019,herviou_multiscale_2019} in our multifractal analysis which is not directly comparable with the (real-space) MBL multifractality proposed by a renormalization group study~\cite{goremykina_analytically_2019}.}}

Our results on multifractality of the MBL phase have the following consequences on other approaches for the MBL problem: (i) they give the correct scaling behavior for variational computational approaches to eigenstates based on the two basis at hand, (ii) put constraints for putative theories of the critical point (in the same spirit as the multifractal spectrum does for AL~\cite{evers_anderson_2008}), (iii) provide {{a new phenomenology, clearly different from the Anderson problem on random graphs or Cayley trees~\cite{altshuler_quasiparticle_1997,
gornyi_interacting_2005,biroli_difference_2012,
de_luca_anderson_2014,altshuler_nonergodic_2016,
tarquini_critical_2017,altshuler_multifractal_2016,
biroli_delocalization_2018,tikhonov_statistics_2019,bera_return_2018,tikhonov_anderson_2016,garcia-mata_scaling_2017}, thus challenging possible analogies.}}

Other perspectives are also opened up by the present work, {{such as connecting with the anomalous slow dynamics observed~\cite{agarwal_anomalous_2015,bar_lev_absence_2015,luitz_extended_2016,khait_transport_2016,steinigeweg_typicality_2016,barisic_dynamical_2016,znidaric_diffusive_2016,luitz_ergodic_2017,bera_density_2017} in a large part of the ETH phase, e.g. 
by considering multifractal properties of operators~\cite{serbyn_thouless_2017}. We note that while rare-regions effects~\cite{agarwal_rare-region_2017} can induce such subdiffusive transport, much less dramatic consequences are expected for eigenstates (reflecting "infinite-time" physics)~\cite{Note_grif}.
Another extension will consist in}} testing for universal behaviour by considering other models with a MBL transition or disturbing the computational basis~\cite{dubertrand_multifractality_2015}. Finally, our results pave the way for a better control of decimation approaches in configuration~\cite{monthus_many-body_2010} or Fock space~\cite{prelovsek_reduced-basis_2018}, as well as for a recently proposed Hilbert space percolation approach~\cite{roy_exact_2018,roy_percolation_2018} to the MBL transition.

\begin{acknowledgments} 
We are deeply grateful to Gabriel Lemari\'e for enlightening explanations about the scaling analysis performed in Ref.~\cite{garcia-mata_scaling_2017}. We also thank D. J. Luitz, Y. Bar Lev, and I. Khaymovich for stimulating discussions.
This work benefited from the support of the project THERMOLOC ANR-16-CE30-0023-02 of the French National Research Agency (ANR) and by the French Programme Investissements d'Avenir under the program ANR-11-IDEX-0002-02, reference ANR-10-LABX-0037-NEXT. We acknowledge PRACE for awarding access to HLRS's Hazel Hen computer based in Stuttgart, Germany under grant number 2016153659, as well as the use of HPC resources from CALMIP (grants 2017-P0677 and 2018-P0677) and GENCI (grant x2018050225).
\end{acknowledgments}

\newpage
$~$
\newpage
\appendix
\setcounter{figure}{0}    
\setcounter{equation}{0}    

\renewcommand{\thefigure}{S\arabic{figure}}
\renewcommand{\theequation}{S\arabic{equation}}

\begin{widetext}

\begin{center}
\vskip 2cm
{\bf{\Large{Supplemental material }}}
\end{center}
\section{ANDERSON PROBLEM IN THE FOCK SPACE}
{\it{(i) Fermionic language---}}
Using the Jordan-Wigner transformation, one can rewrite the spin Hamiltonian 
\be
{\cal H}=\sum_{i=1}^{L}\Bigl[\Delta S_i^z S_{i+1}^{z}-h_iS_i^z+\frac{1}{2}\left(S_i^+ S_{i+1}^{-}+S_i^- S_{i+1}^{+}\right)\Bigr],
\label{eq:Hs}
\ee
as interacting fermions on a chain of $L$ sites
\be
{\cal H}=\sum_{i=1}^{L}\frac{1}{2}\left(f_i^\dagger f_{i+1}^{\vphantom{\dagger}}+{\rm{h.c.}}+2\Delta n_i n_{i+1}-\mu_i n_i\right) +{\rm{C}},
\label{eq:Hf}
\ee
where the random chemical potential $\mu_i=2h_i$, and C is an irrelevant constant. 
The non-interacting part ($\Delta=0$)
\be
{\cal H}_{0}=\sum_{i=1}^{L}\left(f^{\dagger}_{i}f_{i+1}^{\vphantom{\dagger}}+f^{\dagger}_{i+1}f_{i}^{\vphantom{\dagger}}  -\mu_i n _i\right)
\ee 
is diagonalized by $b_k=\sum_{i=1}^{L}\phi_k(i)f_i$ ($\phi_k$ being the single particle orbitals, Anderson localized for any finite disorder in 1D), yielding
%
$
{\cal H}_{0}=\sum_{k=1}^{L}\epsilon_k b^{\dagger}_{k}b_{k}^{\vphantom{\dagger}}$.
In this non-interacting basis, the interaction term $V=\Delta\sum_{i}n_{i} n_{i+1}$ reads
\be
V=\sum_{jklm}V_{jklm}b^{\dagger}_{j}b^{\vphantom\dagger}_{k}b_{l}^{{\dagger}}b_{m}^{\vphantom{\dagger}},\quad {\rm{with}}\quad V_{jklm}=\Delta\sum_{i=1}^{L}\phi_j^*(i)\phi_k(i)\phi_l^*(i+1)\phi_m(i+1).
\label{eq:V}
\ee

{\it{(ii) Many-body problem---}}
When the system is filled with a finite density of fermions $\nu=N_f/L$, many-body states are expressed in the the Fock basis $\{\alpha\}$ 

\be
|{\mathbf{\alpha}}\rangle=\prod_{k=1}^{L}\left(b_k^\dagger\right)^{n_k}|0\rangle\quad (n_k=0,1), \quad {\rm{with}}\quad \sum_{k=1}^{L}n_k=N_f.
\ee
We restrict the discussion to half-filling $\nu=1/2$, putting $N_f=L/2$ particles on $L$ sites yields an Hilbert space of dimension ${\cal N}=L!/(L/2!)^2$. 
The original many-body problem Eq.~\eqref{eq:Hf} can be rewritten in this Fock basis as follows
\be
{\cal H}=\sum_{\alpha=1}^{\cal N}\mu_\alpha|\alpha\rangle\langle\alpha|+\sum_{\alpha\neq\beta}t_{\alpha\beta}|\alpha\rangle\langle\beta|+{\rm{h.c.}},\quad{\rm with}\quad t_{\alpha\beta}=\langle\alpha|V|\beta\rangle\quad {\rm and}\quad
\mu_\alpha=\sum_{k=1}^{L}n_k\epsilon_k+t_{\alpha\alpha}.
\ee
The hopping between sites $\alpha$ and $\beta$ is therefore  governed by the interaction term Eq.~\eqref{eq:V} which can be decomposed in two distinct contributions, 
one-fermion~terms: $V_{jjlm}n_j b_{l}^{{\dagger}}b_{m}^{\vphantom{\dagger}}$ with $L^2/4$ non-zero matrix elements, and two-fermion~terms:
$V_{jklm} b^{\dagger}_{j}b^{\vphantom\dagger}_{k}b_{l}^{{\dagger}}b_{m}^{\vphantom{\dagger}}$ having $L^4/64-L^3/16+L^2/16$ non-vanishing terms.\\

Clearly, off-diagonal hoppings $t_{\alpha,\beta}$ are random numbers, built from products of localized orbitals in Eq.~\eqref{eq:V}.

\section{ALTERNATIVE SCALINGS}
We show in Fig.~\ref{fig:bad} that neither a linear scaling ${\cal{G}}_{\rm lin}\left(\frac{\ln\cal N}{\xi}\right)$ can describe the ETH phase, see Fig~\ref{fig:bad} (a), nor a volumic one ${\cal{G}}_{\rm vol}\left(\frac{\cal N}{\Lambda}\right)$ for the MBL regime, see Fig~\ref{fig:bad} (b). Indeed, we observe systematic deviations as it is impossible to achieve a good collapse of the data, in contrast with Fig.~4 of the main text. Similar scaling violations observed in Fig.~\ref{fig:bad} for ${\overline{S_{2}^{\rm Fock}}}$ occur in the configuration space.
\begin{figure}[h!]
\includegraphics[clip,width=\columnwidth]{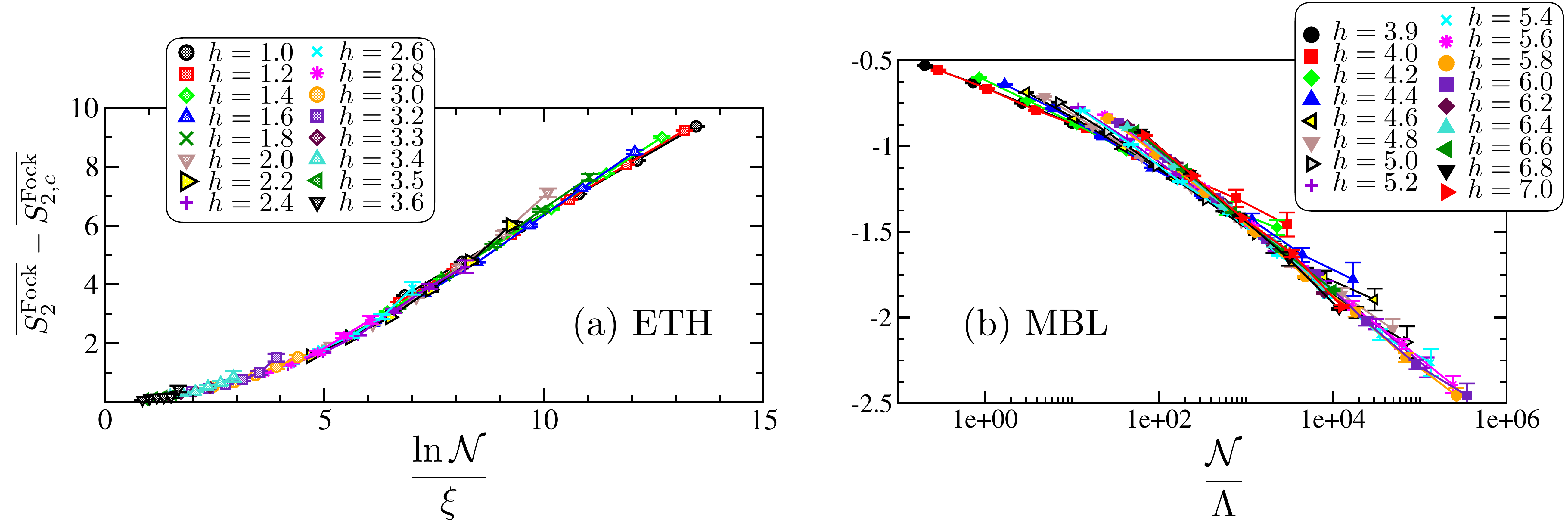}
\caption{Tentative scaling of $q=2$ data in the Fock space (a) in the ETH phase as a function of $\frac{\ln {\cal N}}{\xi}$ and (b) in the MBL regime as a function of ${\cal N}/\Lambda$. The collapse of the data shows systematic deviations, and is clearly less satisfactory as compared to the scaling shown in Fig.~4 of the main text.}
\label{fig:bad}
\end{figure}
$~$\\
\\
\section{CRITICAL FIELD DETERMINATION \& COLLAPSE QUALITY}
Here we address the sensitivity of the scaling analysis with respect the critical disorder value. In our scaling analysis, we used $h_c=3.8$, first guided by the crossing plot (Fig.~3 in the main text). If for instance one takes $h_c=4.4$ as suggested for instance by the numerical linked cluster expansion result~\cite{devakul}, one gets a very poor collapse, as shown below in Fig.~\ref{fig:bad}.
\begin{figure}[h]
\includegraphics[clip,width=.7\columnwidth]{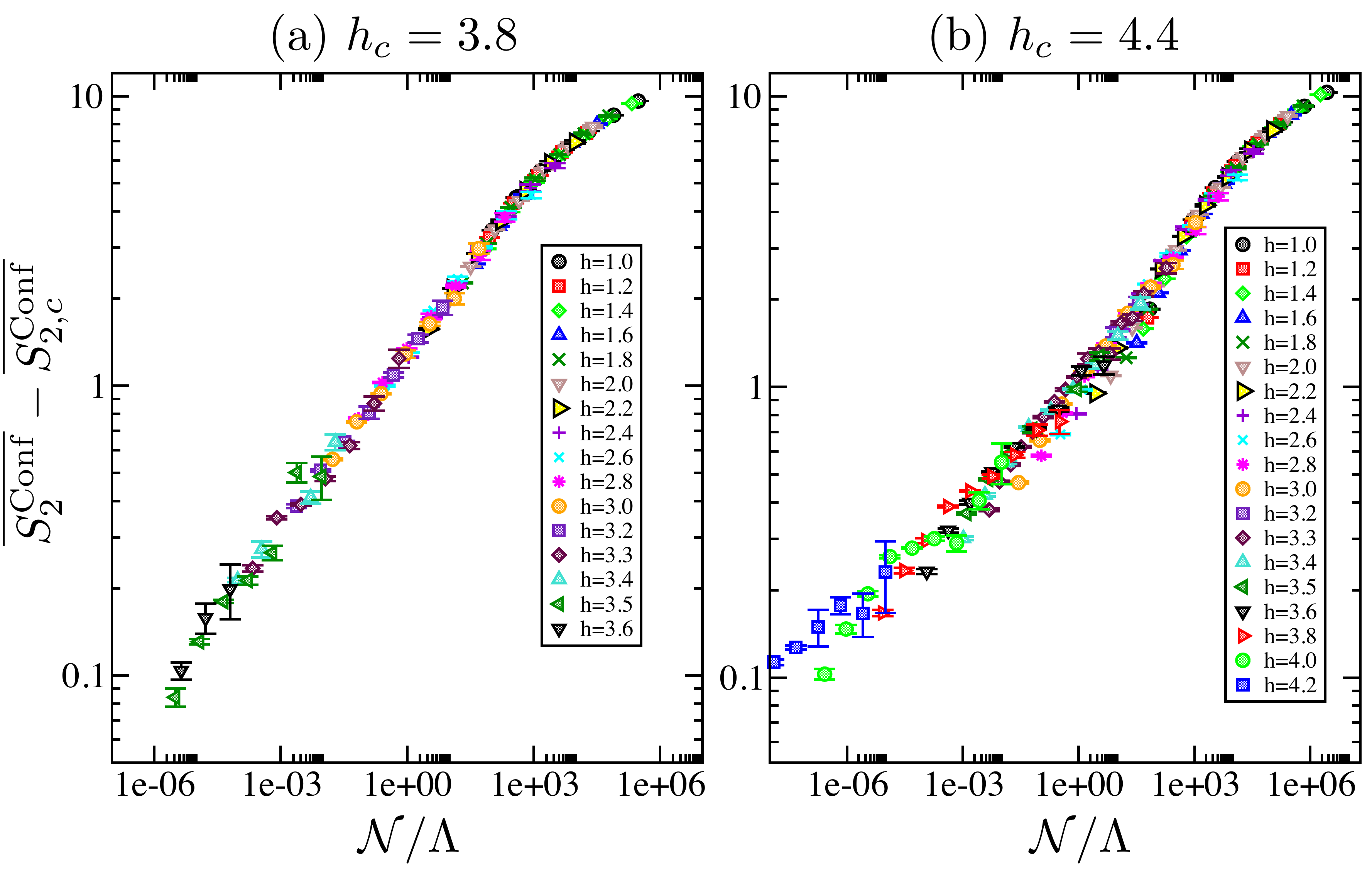}
\caption{Data collapse for the ergodic regime. Comparison between $h_c=3.8$ (a) as shown in Fig.~4 of the manuscript, and $h_c=4.4$ (b) where the collapse is clearly not correct.}
\label{fig:bad}
\end{figure}

One can go further, by making quantitative the quality of the data collapse, which strongly depends on the critical field $h_c$.
Such a determination is non-trivial: while the most commonly accepted value is $h_c = 3.7(2)$ \cite{luitz_many-body_2015}, some works put forward a significantly larger critical disorder \cite{devakul, doggen}. Here, we estimate $h_c = 3.8(1)$, using a method based on the optimization of Spearman's rank correlation coefficient.

Given a 2D cloud of data points $(x_i, y_i)$, Spearman's coefficient $R$ quantifies how close the $y_i$ are from being a monotonic function of the $x_i$. In other words, $R$ quantifies how well the data points align on the graph of an (unspecified) monotonic function.
$R = 1$ indicates a perfectly monotonic relashionship between $y_i$ and $x_i$, while $R = 0$ indicates its complete absence.
In the ETH region, finding the non-ergodicity volumes $\Lambda(h)$ such that $S_q - S_{q,c}$ is best described by the volumic scaling function $\mathcal{G}_\text{vol}\left(\frac{\mathcal{N}}{\Lambda}\right)$ amounts to maximizing Spearman's coefficient $R$.
The residual $1-R$ quantifies the quality of the collapse.
Similarly, in the MBL region, one can maximize Spearman's $R$ to find the lengths $\xi(h)$ such that $S_q - S_{q,c}$ is best described by the linear scaling function $\mathcal{G}_\text{lin}\left(\frac{\log \mathcal{N}}{\xi}\right)$.
This method both automatically determines the collapse and robustly estimates its quality, without any assumption on the form of the scaling function $\mathcal{G}_\text{vol}$ or $\mathcal{G}_\text{lin}$.
Fig.\ \ref{fig:collapse_quality} shows the value of the residual for different candidate critical field values $h_c$.
The best collapse is acheived when the candidate critical field is the closest to the transition point.
The sharp residual dip observed on Fig.\ \ref{fig:collapse_quality} robustly points to a transition at $h_c = 3.8(1)$.
\begin{figure}[h!]
\includegraphics[clip,width=0.7\columnwidth]{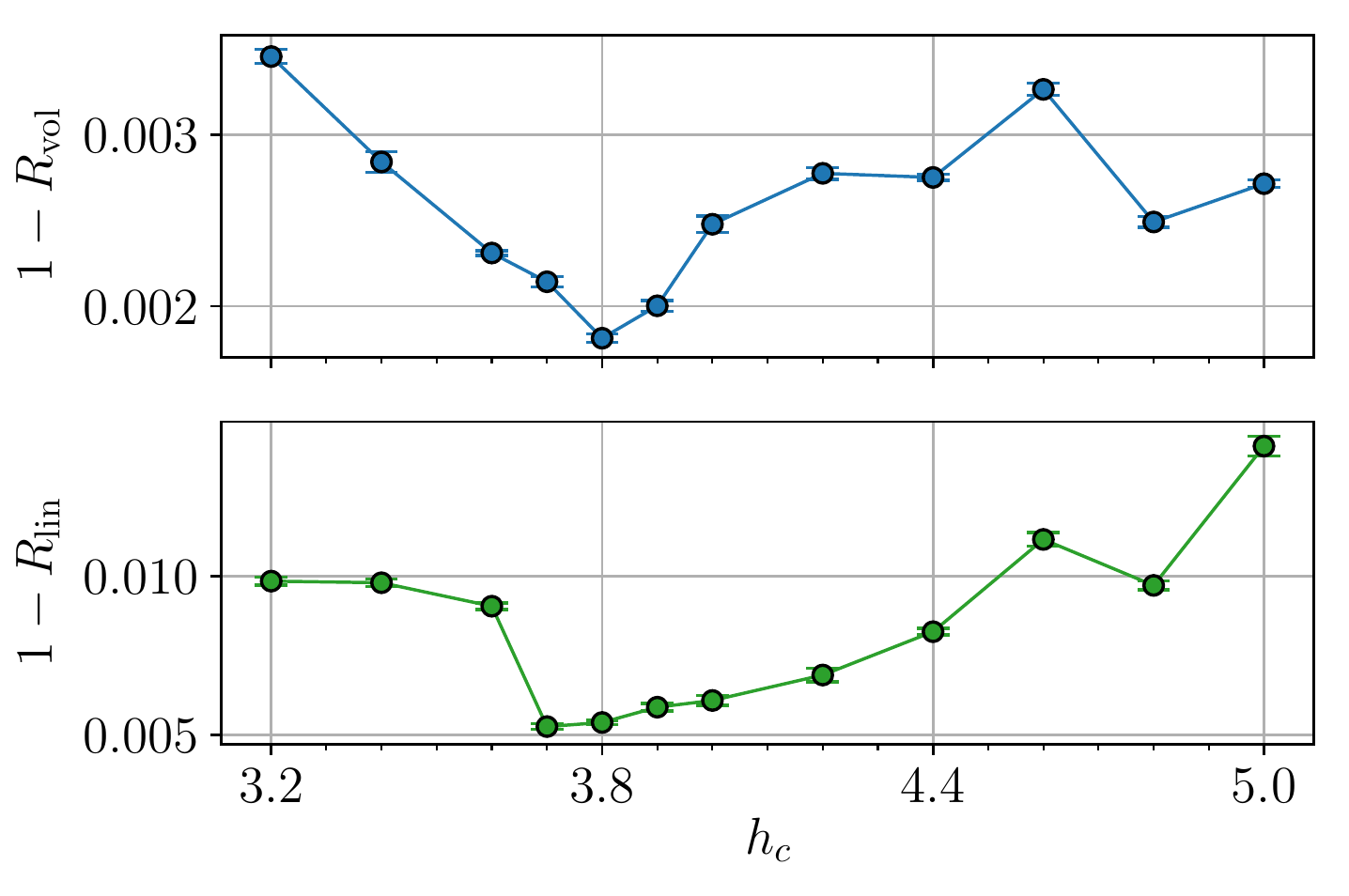}
\caption{Quality (the closest to zero, the better) of the data collapse as a function of the candidate critical disorder $h_c$, using Spearman's rank correlation coefficient $R$. Top panel: quality of a volumic scaling function collapse, using the participations for $h < h_c$. Bottom panel: quality of a linear scaling function collapse, using the participations for $h > h_c$. The quality is evaluated for the configuration basis data.}
\label{fig:collapse_quality}
\end{figure}

$~$\\
\\
\section{DISTRIBUTION OF $S_2$ IN THE MBL REGIME}
The behavior of rare events tails in the MBL regime is an interesting issue for which we provide a more detailed study  at $h=5$, shown below in Fig.~\ref{fig:distribution}, where we confirm that the finite size effects shrink with increasing system sizes. More precisely, the standard deviation $\sigma$ decays with $\ln\cal N$ as a power-law with an exponent $\approx 0.45$, as visible in the inset of Fig.~\ref{fig:distribution}. 
\begin{figure}[h]
\includegraphics[clip,width=.65\columnwidth]{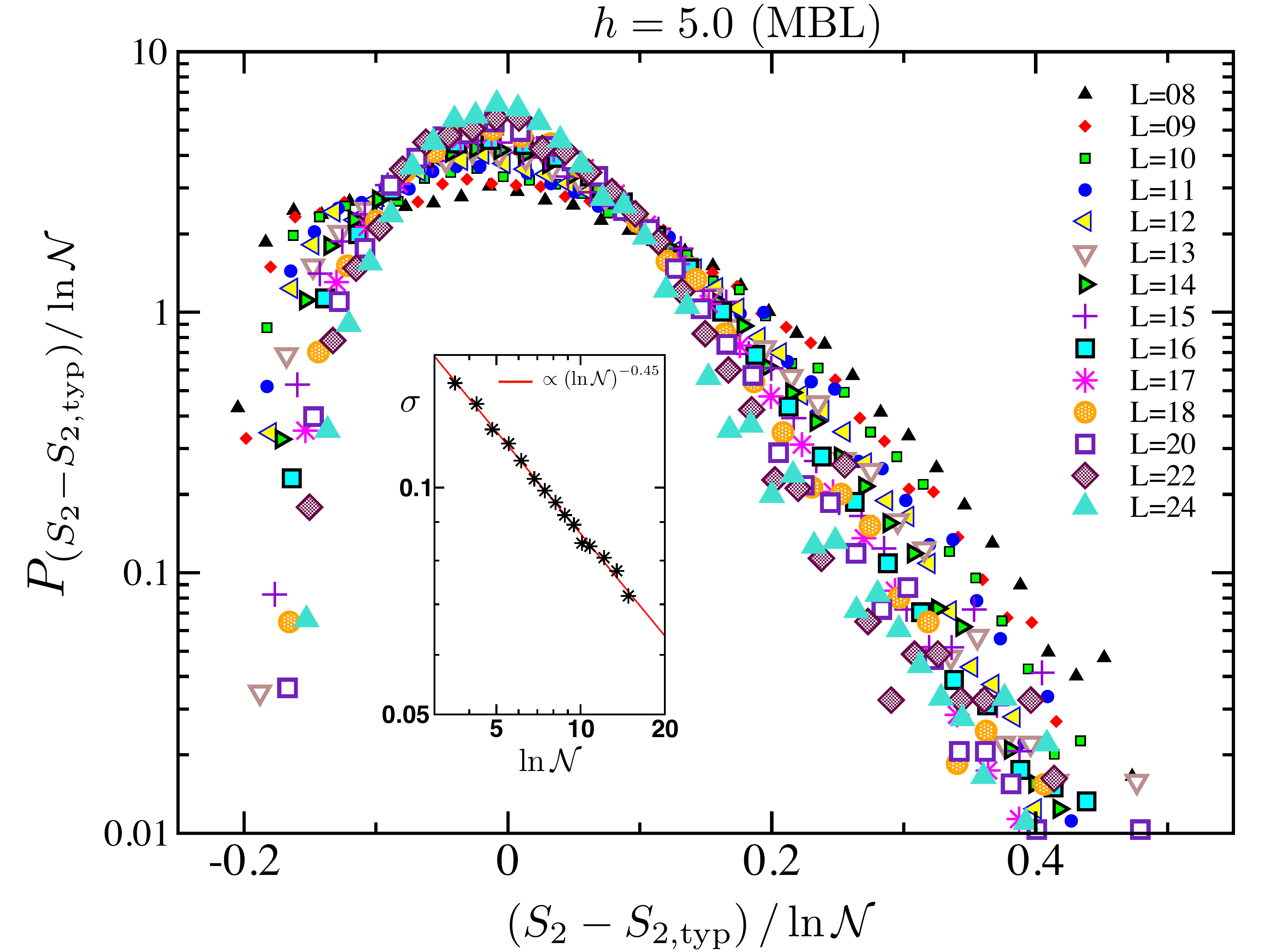}
\caption{Histogram of the participation entropies (normalized by $\ln\cal N$)  rescaled by their typical value. Self-averaging is clearly visible with rare event tails which disappear with increasing system size. In the inset the standard deviation $\sigma$ of the distribution, shown against $\ln\cal N$, decreases as a power-law with an exponent $\approx 0.45$.}
\label{fig:distribution}
\end{figure}

\newpage
$~$\\
\\
\section{STRONG DISORDER LIMIT}
For very high disorder one can use a strong disorder decimation scheme for the spin $S=1/2$ Heisenberg chain
\be
{\cal H}=\sum_{i=1}^{L}\left(J\vec{S}_i\cdot\vec{S}_{i+1}-h_iS_i^z\right).
\label{eq:H}
\ee
The idea is to gradually eliminate sites having the strongest fields.
These sites will be frozen parallel (or anti-parallel) to the field direction for the ground-state (excited state). 
If $|h_i|=h_{\rm max}$, the surrounding sites $(i-1,\,i+1)$ will be perturbatively  coupled by $J_{(i-1,\,i+1)}^{\rm eff}=\frac{J^2}{2h_{\rm max}} \ll J$.
The fields are also weakly renormalized, but we can neglect this small effect for the present argumentation.
The decimation of sites can then be continued up to a cutoff $\Gamma\le |h_i|$, as illustrated in Fig.~\ref{fig:SDRG}.
\begin{figure}[h!]
\includegraphics[clip,width=.85\columnwidth]{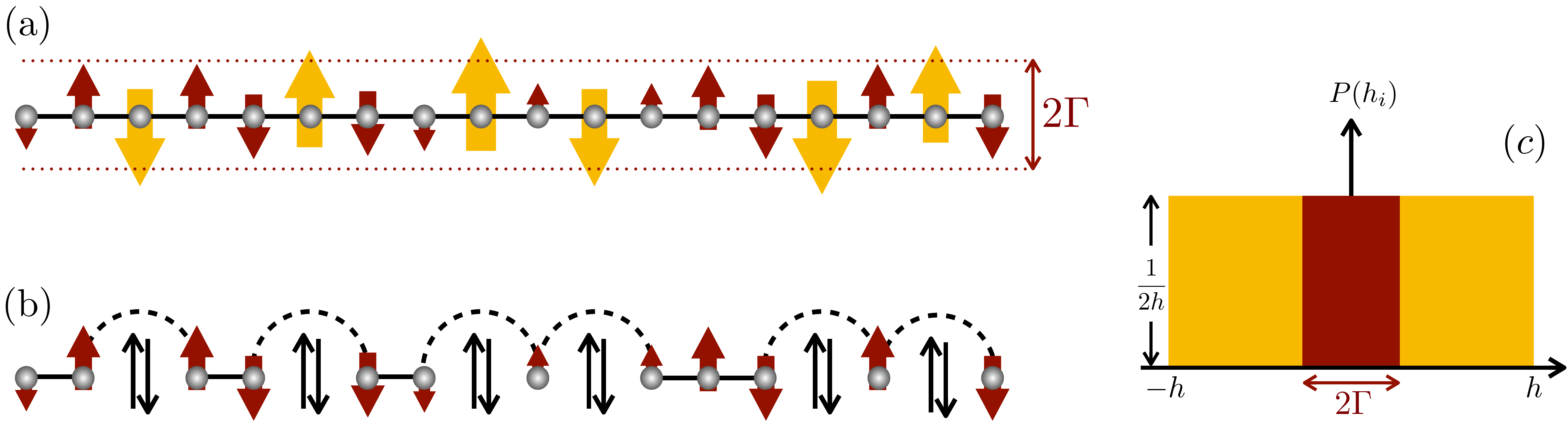}
\caption{Schematic picture for the strong fields decimation process.The sites with strong fields $|h_i|>\Gamma$ shown by yellow arrows in (a) are frozen and the resulting chain is shown in (b) with a reduced number of active sites. Active clusters are connected by small effective couplings $J^{\rm eff}$ (dashed arcs). The uniform distribution cut by $\Gamma$ is illustrated in panel (c).}
\label{fig:SDRG}
\end{figure}

Taking an initial random field distribution uniform in $[-h,h]$ [Fig.~\ref{fig:SDRG} (c)] with $h\gg J$, if we decimate all the fields $|h_i|\ge\Gamma$, a fraction $\rho_{\rm fr}(\Gamma,h)$ of sites will be frozen, while $\rho_{\rm act}=1-\rho_{\rm fr}$ remains active, with $\rho_{\rm act}(\Gamma,h)=\frac{\Gamma}{h}$.
Assuming an initial spin chain with $L$ sites, we end up with an active backbone containing  $L_{\rm act}=L\frac{\Gamma}{h}$ active spins, as shown in Fig.~\ref{fig:SDRG} (b). Only active sites contribute to the participation entropies such that the relevant portion of the spin configuration space will be ${\cal{N}}_{\rm act}\propto {\cal{N}}^{\Gamma/h}$, thus giving to leading order
\be
S_{q}^{\rm Conf}\propto\frac{\Gamma}{h}\ln{\cal N}.
\ee

Ignoring the small couplings $J^{\rm eff}$, the decimated chain acts as a collection of isolated active clusters of various lengths $\ell_{\rm act}^{i}$, and the average size of such active clusters is $\overline{\ell} = {\rho_{\rm fr}}^{-1}=\left({1-\Gamma/h}\right)^{-1}$. Including this length scale in the participation entropy at strong disorder yields
\be
S_{q}^{\rm Conf}\propto\ \left(1-\frac{1}{\overline{\ell}}\right)\ln{\cal N}.
\ee
Moreover, we know from the scaling arguments $S_{q,{\rm MBL}}-S_{q,c}=-D_{q,c}\frac{\ln {\cal N}}{\xi}$ (main text) that in the MBL regime,
\be
S_{q}^{\rm Conf} = D_{q,c} \left(1-\frac{1}{\xi}\right)\ln{\cal N},
\ee
suggesting that at strong disorder, the length scale $\xi$ coincides with the average size of  active clusters: $\xi = \overline{\ell}$.\\

The $1/h$ prediction of this strong disorder argument is well corroborated by the numerics in Fig.~\ref{fig:Sconf_SDRG}. It further confirms that eigenstates cannot be Anderson localized in the spin configuration space, and are generically non-ergodic with a finite multifractal dimension. \\

This naive approach is not expected to work when approaching criticality, for instance it would predict a power-law divergence of $\xi$ with an exponent $a'=1$. We nevertheless find that the prefactor of the $1/h$ decay is such that $S_{q}^{\rm Conf} \approx D_{q,c}\frac{h_c}{h}\ln{\cal N}$ in a fair range of large $h$, a formula which is used to plot the straight lines in Fig.~\ref{fig:Sconf_SDRG}. \\

\begin{figure}[htp]
\includegraphics[clip,width=.82\columnwidth]{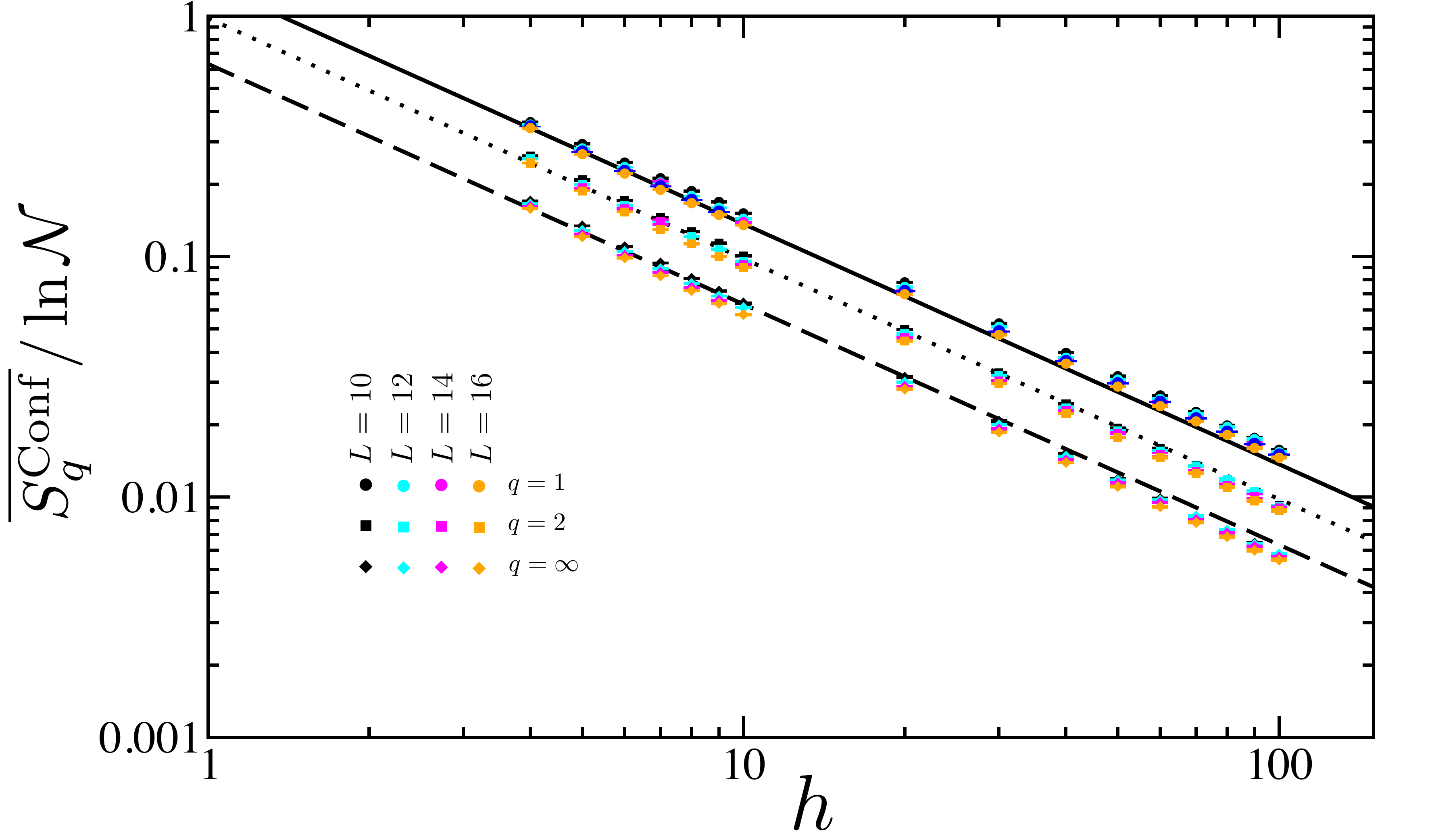}
\caption{Comparison of ED results on small system sizes at very strong disorder (using about several tens of thousands disorder realizations) with the expression $S_{q}^{\rm Conf} = D_{q,c}\frac{h_c}{h}\ln{\cal N}$ for $q=1$ (full line), $q=2$ (dotted line), and $q=\infty$ (dashed line). Here we use $D_{q,c}$ as obtained from the $L=16$ data.}
\label{fig:Sconf_SDRG}
\end{figure}


\begin{thebibliography}{81}
\bibitem [{not({\natexlab{a}})}]{noteED}%
  \BibitemOpen
  \href@noop {} {} \bibinfo {note} {Note that while a direct
  ED of the hopping problem Eq.~\eqref{eq:Hc} is possible up to $L=24$ sites in
  the configuration basis, this is not practically possible in the Fock space
  which suffers from a much higher connectivity. We have therefore privileged
  to perform rotations form configuration to Fock basis, still computationally
  costly but allowing us to get PE data up to $L=22$ in the Fock
  space.}\BibitemShut {Stop}%
\bibitem [{not({\natexlab{b}})}]{note_collapse}%
  \BibitemOpen
  \href@noop {} {} \bibinfo {note} {In~\cite{sm} we show
  other tentative scalings, clearly less satisfactory.}\BibitemShut {Stop}%
\bibitem [{Not()}]{Note_hc}%
  \BibitemOpen
  \href@noop {} {}\bibinfo {note} {The choice of $h_c=3.8$ is very carefully
  justified by a quantitative estimate of the sensitivity in the value of
  $h_c$, see \cite{sm}.}\BibitemShut {Stop}%
\bibitem [{not()}]{note_PL}%
  \BibitemOpen
  \href@noop {} {}\bibinfo {note} {We cannot entirely rule out a power-law
  divergence for $\Lambda$, although it would yield a very high exponent
  $\approx 7$.}\BibitemShut {Stop}%
\bibitem [{Not()}]{Note_grif}%
  \BibitemOpen
  \href@noop {} {}\bibinfo {note} {Consider for example a site with an
  anomalously large local field: this tends to polarize the corresponding spin
  and leads to a bottleneck, impeding transport. However, it will not change
  the volumic delocalization exponent $D_q=1$, but can only modify subleading
  terms, at most $-\ln 2$ in the extreme case of full
  polarization.}\BibitemShut {Stop}%
\end{thebibliography}
\end{widetext}
\end{document}